# Optical identification of hybrid magnetic and electric excitations in $Dy_3Fe_5O_{12}$ garnet

P. D. Rogers,[1a] Y. J. Choi,[2] E. Standard,[1] T. D. Kang,[1] K. H. Ahn,[1] A. Dubroka,[3] P. Marsik,[3] Ch. Wang,[3] C. Bernhard,[3] S. Park,[2] S-W. Cheong,[2] M. Kotelyanskii,[1,4] and A. A. Sirenko[1]

[1] *Department of Physics, New Jersey Institute of Technology, Newark, New Jersey 07102, USA*

[2] *Rutgers Center for Emergent Materials and Department of Physics and Astronomy, Rutgers University, Piscataway, New Jersey 08854, USA*

[3] *Department of Physics, University of Fribourg, CH-1700 Fribourg, Switzerland*

[4] *Rudolph Technologies Inc., Flanders, New Jersey 07836, USA*

---

[a] pdr2@njit.edu

# ABSTRACT

Far-infrared spectra of magneto-dielectric $Dy_3Fe_5O_{12}$ garnet were studied between 13 and 100 cm$^{-1}$ and at low temperatures between 5 and 80 K. A combination of transmission, reflectivity, and rotating analyzer ellipsometry was used to unambiguously identify the type of the dipole activity of the infrared modes. In addition to purely dielectric and magnetic modes, we observed several hybrid modes with a mixed magnetic and electric dipole activity. These modes originate from the superexchange between magnetic moments of Fe and Dy ions. Using $4 \times 4$ matrix formalism for materials with $\mu(\omega) \neq 1$, we modeled the experimental optical spectra and determined the far-infrared dielectric and magnetic permeability functions. The matching condition $\mu(\omega_h) \cdot S_e = \varepsilon(\omega_h) \cdot S_m$ for the oscillator strengths $S_{e(m)}$ explains the observed vanishing of certain hybrid modes at $\omega_h$ in reflectivity.

Far-infrared (IR) spectra of the optical modes in magnetic materials have recently attracted a lot of attention, especially with respect to the multiferroic effect and electromagnons [1-5]. The theory of coupling between magnetic and electric excitations is not well developed yet, but understanding can be improved with the help of experimental measurements of both the complex dielectric function $\varepsilon(\omega)$ and the magnetic permeability $\mu(\omega)$ in the frequency range of the magnetic excitations. One possible experimental solution to this problem is based on the combination of optical measurements in transmission and reflection geometry [6]. The quantitative interpretation of the optical spectra requires an adequate modeling approach for light propagation in magneto-dielectric crystals with $\mu(\omega) \neq 1$. In this paper, we apply Berreman's 4×4 matrix formalism [7] for the numerical and analytic analysis of experimental data for transmission, reflectivity, and rotating analyzer ellipsometry (RAE) in single crystals of Dysprosium Iron garnet (Dy-IG). Through the combination of these optical techniques, we show that one can determine whether an IR-active mode is (i) entirely of dielectric origin, (ii) entirely of magnetic origin, or (iii) a hybrid with a mixed electric- and magnetic dipole activity. The observed vanishing of certain hybrid modes in the reflectivity spectra of Dy-IG is explained in terms of the adjusted oscillator strength matching (AOSM), derived using 4×4 matrix formalism.

The high-temperature flux growth technique was utilized to produce bulk crystals of Dy-IG ($Dy_3Fe_5O_{12}$). A sample with a (0 0 1) surface, a cross section area of 5×5 mm$^2$, thickness of 0.55 mm, and a 3° offset between opposite sides was used for the optical experiments. Transmission spectra with resolution of 0.3 cm$^{-1}$ were measured between 13 and 100 cm$^{-1}$ at the National Synchrotron Light Source, Brookhaven National Laboratory, at the U4IR beamline equipped with a Bruker IR spectrometer, and a LHe-pumped bolometer. The RAE and reflectivity measurements were carried out at Fribourg University using a Hg lamp in the spectral range between 45 and 100 cm$^{-1}$ with resolution of 0.7 cm$^{-1}$. The RAE experimental setup is similar to that described in Ref. [8]. Temperature and magnetic field dependencies for static values of $\varepsilon(0,H,T)$ and $\chi(0,H,T)$ were measured using an LCR meter at 44 kHz and a SQUID magnetometer.

Dy-IG, as well as other $RE$-IG ($RE$=Ho, Tb), is a ferrimagnetic material with a huge magnetostriction, which is related to the combination of a strong anisotropy of the crystal field of the $RE^{3+}$ ions and a strong and anisotropic superexchange interaction between $RE$ and iron [9-12]. Below the transition temperature of $T_N \approx 550$ K, the iron spins are ordered in a ferrimagnetic structure along the [1 1 1] direction. At low temperatures, the Dy spins are antiparallel to the net magnetic moment of Fe. While the iron sublattice magnetization does not change appreciably below 100 K, the Dy sublattice magnetization increases rapidly towards low temperature. Below 100 K, a rhombohedral distortion of the cubic cell causes the canting of Dy spins, which form a double umbrella structure [13]. Although there are no literature reports that Dy-IG is multiferroic, recently two related compounds, antiferromagnetic orthoferrite $DyFeO_3$ and Tb-IG, were shown to be multiferroic and magneto-dielectric [14,15].

In addition to the aforementioned properties of Dy-IG, we also found a magneto-dielectric effect in a weak external magnetic field of about 2 kOe. We observed two indications of the ferromagnetic ordering of Dy spins at $T_C$=16 K: (i) the sharp minimum in the temperature derivative of magnetic susceptibility $\partial\chi/\partial T$ at $T_C$ [Fig. 1(a)] and (ii) the temperature dependence of the exchange resonance IR mode frequencies, which will be discussed below. The quasi-static value of the dielectric constant $\varepsilon(0)$ of Dy-IG has anomalies in the temperature and external magnetic field dependencies [Fig. 1(b,c)]. $\varepsilon(0,T)$ has a peak at $T_C$=16 K that can be explained by the local electric polarization due to anti-ferroelectric lattice ordering. The latter occurs in the same temperature range as the ferromagnetic ordering of the Dy spins below 16 K. The anti-ferroelectric lattice ordering does not create a global electric polarization, but affects the spin and lattice dynamics at the microscopic scale. Recently, local electric polarization has also been observed in Tb-IG [16]. Using RAE we found that the soft optical modes at $\Omega \approx$ 146 cm$^{-1}$ and 595 cm$^{-1}$, which are associated primarily with Dy and oxygen displacements, contribute to the changes in $\varepsilon(0,T)$ through the Lyddane-Sachs-Teller relationship $\varepsilon(0,T) \sim \Omega^{-2}(T)$ [see Figure 1(b)]. The magneto-dielectric effect in Dy-IG reveals itself in the variation of $\varepsilon(0,H)$ for $H$<10 kOe [Fig. 1(c)], which can be qualitatively explained by a suppression of the anti-ferroelectric order by the weak external magnetic field $H$.

The appearance of anti-ferroelectric ordering and a Dy–Dy ferromagnetic interaction motivates us to re-visit the far-IR optical spectra of Dy-IG. *RE*-IGs have been studied in Refs. [17-20]. It was shown that below 80 cm$^{-1}$, transmission spectra in polycrystalline *RE*-IGs are dominated by both $RE^{3+}$ single ion electronic transitions and Kaplan-Kittel (KK) modes, which were attributed to magnetic dipoles [21]. Figure 2(a,b) shows a transmission spectrum of Dy$_3$Fe$_5$O$_{12}$ at $T$ = 5 K, and the transmission intensity map measured with the temperature increments of 1 K. In addition to the optical phonon at 81 cm$^{-1}$ [22], a number of lines at 20, 52, 72, and 87 cm$^{-1}$ is observed for $T$ > 16 K. Their temperature-independent behavior is typical for the crystal field (CF) transitions of Dy$^{3+}$. At low temperatures $T$<16 K, however, the number of absorption lines increases compared to that at higher temperatures [Fig. 2(b)]. The ligand field (LF) and KK modes appear at 13, 22, 29, 43, 51, 59.5, 73, 78, 87, 91, and 98 cm$^{-1}$ for $T$=5 K. As shown below, the LF and KK modes can be distinguished based on the temperature dependence of their frequencies. In a simplified model for two-spin ferrimagnetic systems, like *RE*-Fe, a single exchange-type KK mode is expected with the frequency of $\omega_M$. The LF mode $\omega_{LF}$ corresponds to precession of the Dy$^{3+}$ moments in the effective field imposed by the iron magnetization due to the superexchange interaction between Fe and *RE*. The latter is modified by the ferromagnetic interaction between Dy$^{3+}$ spins at low temperature. The zone-center frequencies of these collective excitations of Dy$^{3+}$ and Fe$^{3+}$ spins are [17,18,21]:

$$\omega_M(T) = \lambda_{Fe-Dy}\mu_B \left[ g_{Dy}M_{Fe} - g_{Fe}M_{Dy}(T) \right]$$
$$\omega_{LF}(T) = g_{Dy}\mu_B \left[ \lambda_{Fe-Dy}M_{Fe} + \lambda_{Dy-Dy}M_{Dy}(T) \right],$$
(1)

where $\mu_B$ is the Bohr magneton, $\lambda_{Fe-Dy}$ is the exchange constant between Fe and Dy ions, $\lambda_{Dy-Dy}$ is the ferromagnetic exchange constant, $g_{Fe} = 2$ and $g_{Dy}$ are the corresponding g-factors, $M_{Dy}(T)$ is the Dy-sublattice magnetization, and $M_{Fe}$ is the combined Fe magnetization. For $T < 16$ K, the KK modes $\omega_M(T)$ exhibit softening of the frequency due to increase of the Dy magnetization $M_{Dy}(T)$. Figure 2(a,b) shows three KK modes at 43, 51, and 59.5 cm$^{-1}$, that can be explained by the double umbrella structure for Dy$^{3+}$ spins and by the strongly anisotropic and temperature-dependent superexchange interaction between Dy$^{3+}$ and Fe$^{3+}$ ions. The temperature-induced variation of the LF mode frequencies below 16 K is also proportional to $M_{Dy}(T)$ [see Eq.(1)], but it has an opposite sign compared to that for KK modes. Fig. 2(b) indicates a phase transition at $T_C$ =16 K with appearance of the long range ordering of Dy spins.

According to the simplified model for collinear Dy$^{3+}$ and Fe$^{3+}$ spins, the KK and LF modes can be viewed as pure magnons which contribute only to the magnetic permeability $\mu(\omega)$ [18,21]. However, their spectral proximity to the phonon at 81 cm$^{-1}$ and modification of the LF due to local electric polarization should result in a hybrid electric- and magnetic-dipole activity. In the following, we will prove this suggestion using a combination of several optical techniques: transmission and reflectivity at normal incidence, and RAE. The terms "LF" and "hybrid" will be applied interchangeably to the same modes. The first term refers to the origin of the IR-active excitation as described above, while the latter corresponds to the mixed dipole activity of the mode in the optical spectra.

Figure 3(a,b) compares the transmission $T_s(\omega)$ and reflectivity $R_s(\omega)$ spectra of the same Dy-IG sample as in Figure 2. $T_s(\omega)$ and $R_s(\omega)$ have been measured at $T=8$ K and 9 K, respectively, at near-normal incidence, *i. e.*, the angle of incidence (AOI) is close to zero. RAE measurements were taken for the same sample at $T=8$ K and AOI=75 deg. The results of the RAE measurements are shown in terms of the real part of the pseudo-dielectric function $\langle \varepsilon_1(\omega) \rangle$, [Fig. 3(c)]. Modes of three kinds can be identified in Fig. 3(a,b,c): (i) The phonon at 81 cm$^{-1}$, which is obviously an electric dipole, has a conventional Lorentz shape in the $R_s(\omega)$ and RAE spectra. The phonon is also strong in $T_s(\omega)$; (ii) The KK mode at 59.5 cm$^{-1}$ has an inverted Lorentz shape in both the $R_s(\omega)$ and RAE spectra. As shown below, this shape is typical for magnetic dipoles. (iii) The LF modes at 73, 78, and 91 cm$^{-1}$ are as strong as the phonon in $T_s(\omega)$, but practically invisible in both the $R_s(\omega)$ and RAE spectra. These complementary results in the $T_s(\omega)$ and $R_s(\omega)$ spectra, both measured for the same sample and at the same AOI, can be reconciled by suggesting that the LF modes in Dy-IG possess a hybrid, *i.e.*, magnetic- and electric-dipole activity.

The inverted Lorentz shape for a magnetic dipole, the possible suppression of the hybrid modes in the $R_s(\omega)$ and RAE spectra, and the strong contribution of these modes in the transmission spectra can all be qualitatively understood based on Veselago's approach for light propagation in an isotropic, semi-infinite medium with $\mu(\omega) \neq 1$. Here a simple replacement of the refractive index is used: for Fresnel's reflection coefficient, $n(\omega) \to \sqrt{\varepsilon(\omega)/\mu(\omega)}$; while the propagation in the medium and thus the transmission spectra are driven by: $n(\omega) \to \sqrt{\varepsilon(\omega) \cdot \mu(\omega)}$ [23,24]. This explains that a purely magnetic mode has an inverted shape in the reflectivity spectrum since $n(\omega) \sim \sqrt{1/\mu(\omega)}$ in the vicinity of the mode where $\varepsilon(\omega) \approx const$. It also naturally accounts for the suppression of the mode feature in the reflectivity spectrum for a hybrid, *i.e.*, magnetic-dielectric mode, where the magnetic and dielectric components tend to cancel each other, while remaining additive in the transmission spectrum. These qualitative arguments allow for an unambiguous identification of the magnetic, dielectric and/or hybrid nature of the infrared-active modes based on combined reflection and transmission data. Details of these qualitative derivations are summarized in the Supplemental Appendix [25].

In order to properly analyze the experimental data in Fig. 3(a,b,c), we developed an exact numeric method (see Ref. [26] for details), which is based on Berreman's 4×4 matrix formalism [7,27] that can be applied to magnetic materials in both semi-infinite and thin film configuration. In the case of a thin film, such an approach does not require any simplifications or initial assumptions about the sample thickness and the absorption coefficients for each mode. Our method incorporates the exact geometry of the measured Dy-IG sample with average thickness $d$=0.55 mm, multiple reflections, variable AOI's, and possible magnetic and electric anisotropies. The results of our numeric method were used to model the $R_s(\omega)$, $T_s(\omega)$, and RAE spectra and to determine the parameters of the electric, magnetic and/or hybrid dipole activity. Our approach, of course, reproduces exactly Veselago's results for reflection from a semi-infinite, isotropic sample referred to in the qualitative discussion above. The response functions of Dy-IG, $\varepsilon(\omega)$ and $\mu(\omega)$, were modeled using a set of Lorentz oscillators:

$$\varepsilon(\omega) = \varepsilon_\infty + \sum_{j=1}^{N} \frac{S_{j,e}\omega_{j,e0}^2}{\omega_{j,e0}^2 - \omega^2 - i\gamma_{j,e}\omega},$$

$$\mu(\omega) = \mu_\infty + \sum_{j=1}^{M} \frac{S_{j,m}\omega_{j,m0}^2}{\omega_{j,m0}^2 - \omega^2 - i\gamma_{j,m}\omega}.$$

(2)

Here $\varepsilon_\infty$ is the infinite-frequency value of the dielectric function that contains the contribution of the high-frequency phonons and interband electronic excitations, $\mu_\infty \cong 1$, $S_{e(m)}$ is the corresponding mode oscillator strength, $\gamma_{e(m)}$ is the damping constant, and $\omega_{e(m)0}$ is the resonance frequency. Although the response functions of Dy-IG can be in principle anisotropic, the comparison of the reflectivity and ellipsometric data taken at different AOI do not reveal any anisotropy within the accuracy of the data. The hybrid modes in this model have non-zero electric and magnetic oscillator strengths $S_e$ and $S_m$ at the same resonant frequency $\omega_h = \omega_{e(m)0}$, thus creating a contribution to both $\varepsilon(\omega)$ and $\mu(\omega)$. The electric and magnetic damping constants for the hybrid modes are assumed to be the same: $\gamma_e = \gamma_m$. The results of the fit using 4×4 matrix formalism for $R_s(\omega)$, $T_s(\omega)$, and $\langle\varepsilon_1(\omega)\rangle$ are shown in Figs. 3(a,b,c) with solid curves. The corresponding values of $S_e$ and $S_m$ are summarized in Table I and the real parts of the dielectric function and the magnetic permeability are shown in Figure 3(d,e), where the hybrid modes are marked with $h$. Note that for Dy-IG, $S_e$ and $S_m$ are not large enough to modify significantly the background values of $\varepsilon_\infty \cong 17$ and $\mu_{bg} \cong 1$. Hence, both $\varepsilon(\omega)$ and $\mu(\omega)$ are positive everywhere in the vicinity of the hybrid mode frequencies [see in Fig. 3(d,e)]. Thus, the natural occurrence of a negative index of refraction does not take place at the spectral range dominated by the hybrid modes that might otherwise occur if their damping was sufficiently low.

In order to illustrate how the Lorentzian parameters quantitatively influence the measured $T_s(\omega)$, $R_s(\omega)$, and RAE spectra, certain analytical formulas can be obtained. Consider two electric and magnetic oscillators that are well-separated on the energy scale and have comparable values of $\gamma_e \approx \gamma_m$. If the backside reflected beams are not strong, the ratio of the amplitudes of the modes in the reflectivity spectra at their respective resonances are related to $\partial R_{ss}(\omega)/\partial \omega |_{\omega e(m)0}$ as follows:

$$\frac{\partial R_{ss}/\partial \omega |_{\omega e 0}}{\partial R_{ss}/\partial \omega |_{\omega m 0}} \approx -\frac{\mu_\infty}{\varepsilon_\infty} \frac{S_e}{S_m} \frac{\omega_{e0}}{\omega_{m0}}, \quad (3)$$

where $S_e \ll \varepsilon_\infty$. $\mu_\infty$ and $\varepsilon_\infty$ are determined at the frequencies shifted from $\omega_{e(m)0}$ by at least $3\gamma_{e(m)}$. Note that the negative sign corresponds to the inverted Lorentzian shape at the magnetic resonance, as was discussed earlier. If the thickness of the sample $d$ is optimized to prevent saturation of the transmitted intensity at the resonance, then the following relationship for transmission amplitudes of the magnetic- and electric modes can be obtained:

$$\frac{\Delta T_e}{\Delta T_m} \approx \frac{\mu_\infty}{\varepsilon_\infty} \frac{S_e}{S_m} \frac{\omega_{e0}^2}{\omega_{m0}^2} \quad (4)$$

where $\Delta T_{e(m)} \approx T(\omega_{e(m)0}) - T(\omega_{e(m)0} \pm 3\gamma_{e(m)})$. In the case of hybrid modes with a mixed electric- and magnetic dipole activity, Eq. (3) and Eq. (4) indicate that the contribution of the dielectric and magnetic oscillators to the transmission spectra is additive with an adjusted oscillator strength (AOS) $S_T \approx \mu_\infty \cdot S_e + \varepsilon_\infty \cdot S_m$, while their total contribution to reflectivity is subtractive with AOS of $S_R = (\mu_\infty \cdot S_e - \varepsilon_\infty \cdot S_m)/\mu_\infty^2 \approx \mu_\infty \cdot S_e - \varepsilon_\infty \cdot S_m$. Here, the relevant magnetic or dielectric oscillator strength is multiplied by its constitutive response function complement. For the general case of a spectrum with several hybrid modes and a moderate overlap between them, a complete cancellation in reflectivity measurements is possible for each mode if the adjusted oscillator strength matching condition (AOSM) occurs: $\mu(\omega_h) \cdot S_e = \varepsilon(\omega_h) \cdot S_m$. This condition, derived using 4×4 matrix formalism, is consistent with Veselago's qualitative approach when Lorentz oscillators are used to describe the response functions for a single mode: $\mu_\infty \cdot S_e = \varepsilon_\infty \cdot S_m$. For brevity, derivations for AOS and AOSM are contained in Supplemental Appendix [25]. The AOSM condition is realized for the hybrid modes at 73 and 78 cm$^{-1}$ that are not visible in either normal-incidence reflectivity or RAE experiments. The hybrid mode contribution to $dR_{ss}(\omega_h)/d\omega$ is negligible and the $R_s(\omega)$ spectrum looks essentially featureless around the resonance frequencies. The analysis of RAE spectra taken at AOI=75° shows that the AOSM condition $\mu(\omega_h) \cdot S_e = \varepsilon(\omega_h) \cdot S_m$ is valid across a wide range of AOIs, even close to the Brewster angle (76.4° for $\varepsilon_\infty = 17$ and $\mu_\infty = 1$).

In conclusion, the rare occurrence of hybrid modes has been studied in Dy-IG. We speculate that the proximity of the Dy$^{3+}$ LF exchange resonances (73 and 78 cm$^{-1}$) to the frequency of the lowest optical phonon (81 cm$^{-1}$), local electric polarization, and the non-collinear spin structure for the Dy-Fe magnetic system are responsible for the mode hybridization. All these conditions provide the possibility to transfer the oscillator strength from the phonons to the LF resonances. Berreman's 4x4 matrix formalism was used to derive analytical relationships that describe the inverted Lorentz shape for magnetic dipoles in reflectivity and RAE experiments and provides the correct ratio of oscillator strengths for magnetic, electric, and hybrid excitations. The AOSM condition is used to explain the almost complete cancellation of the hybrid modes in the reflectivity spectra while remaining strong in the transmission spectra. The hybrid modes have a potential for a natural realization of the negative index of refraction in magnetic materials in proximity to $\omega_h$. One of the possible applications of the AOSM condition is for the design of antireflective coatings in the far-IR spectral range using magnetic- and metamaterials.

The authors are thankful to T. Zhou for valuable discussions and to G. L. Carr for help at U4IR beamline. The far-IR Transmission experiments at NJIT and the crystal growth at Rutgers University were supported by DOE DE-FG02-07ER46382. Use of the National Synchrotron Light Source, Brookhaven National Laboratory, was supported by the U.S. Department of Energy, Office of Science, Office of Basic Energy Sciences, under Contract No. DE-AC02-98CH10886. The ellipsometry and reflectivity measurements at University of Fribourg were supported by LiMAT Fellowship and by the Schweizerische Nationalfonds (SNF) by grants 200020-129484 and the NCCR-MaNEP. The theory development for the 4×4 matrix formalism was supported by NSF-DMR-0821224.

TABLE I. The values of parameters of optical phonon at 81 cm$^{-1}$ (*e*), magnetic KK mode at 59.5 cm$^{-1}$ (*m*), and three hybrid modes (*h*) at 73 cm$^{-1}$, 78 cm$^{-1}$ and 91 cm$^{-1}$ obtained from the analysis of the combination of the transmission, RAE and reflectivity measurements.

| $\omega_0$, cm$^{-1}$ | $S_e$ units of $\varepsilon$ | $S_m$ units of $\mu$ | Type |
|---|---|---|---|
| 59.5 | – | 0.0019 | *m* |
| 73 | 0.036 | 0.0021 | *h* |
| 78 | 0.035 | 0.0022 | *h* |
| 81 | 0.077 | – | *e* |
| 91 | 0.032 | 0.0010 | *h* |

FIGURES

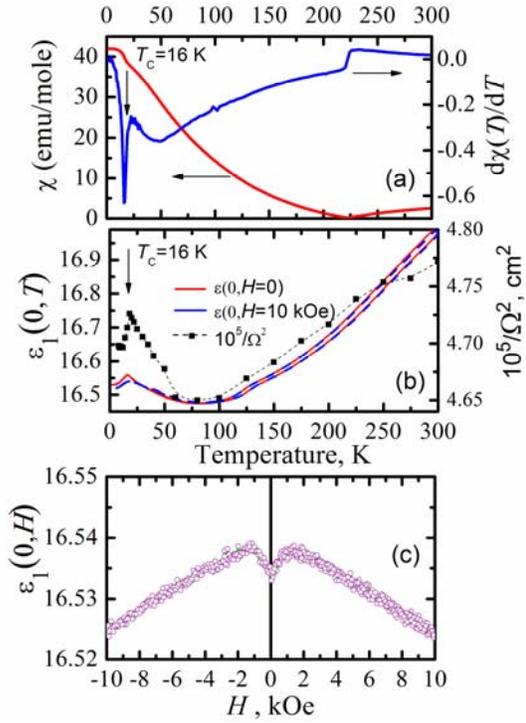

FIG. 1 (color online). (a) Temperature dependence of the static magnetic susceptibility (red curve) and its derivative (blue curve) for a $Dy_3Fe_5O_{12}$ single crystal. Ferromagnetic ordering of $Dy^{3+}$ occurs at $T_C$ =16 K. (b) Temperature dependence of the static dielectric constant at $H$=0 (red) and $H$=10 kOe (blue). Black squares represent the temperature dependence of the soft optical phonon frequency at 146 cm$^{-1}$ measured with RAE. (c) Magnetic field dependence of the static dielectric constant at $T$= 5 K. In all graphs $E \parallel [1\ 0\ 0]$ and $H \parallel [0\ 1\ 1]$.

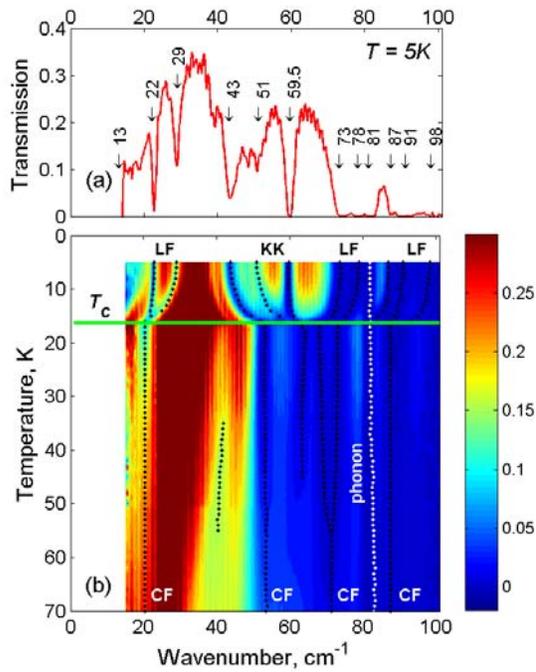

FIG. 2 (color online). (a) Far-IR transmission spectrum for a $Dy_3Fe_5O_{12}$ single crystal measured at $T$=5 K. The light propagation is along the [0 0 1] direction. Arrows indicate the frequencies of the IR modes. (b) Transmission map *vs.* temperature and light frequency. The blue (dark) color corresponds to stronger absorption and red (light) color indicates high transmission. The horizontal green line represents the ferromagnetic transition temperature $T_C$ =16 K. The white dots represent the phonon at 81 cm$^{-1}$. The black dots show the KK and LF excitations.

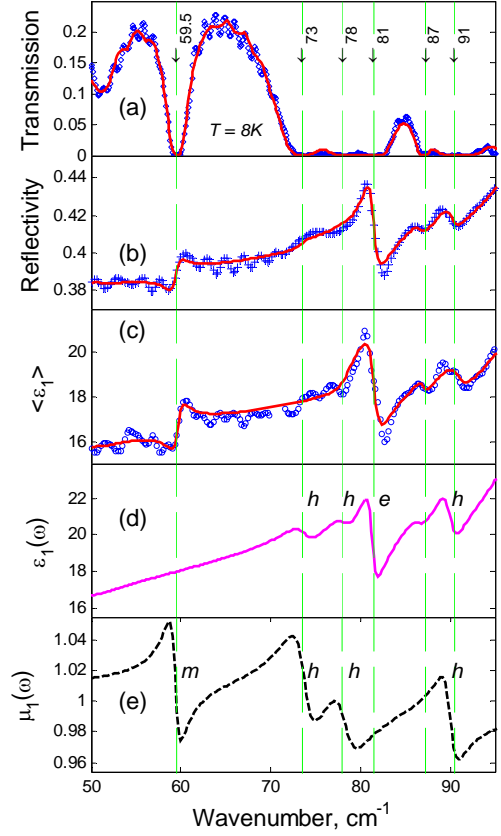

FIG. 3 (color online). Optical spectra of a $Dy_3Fe_5O_{12}$ single crystal. (a) Transmission spectrum at AOI=0, $T$=8 K. (b) Absolute far-IR reflectivity at AOI=0, $T$=9 K. (c) Rotating analyzer ellipsometry (RAE) data for pseudo dielectric function $\langle \varepsilon_1(\omega) \rangle$ at AOI=75 deg, $T$=8 K. Red solid curves in (a,b,c) represent results of the fit. Electric (d) and magnetic (e) susceptibilities as determined from the fit results. Magnetic, electric, and hybrid modes are marked with *m*, *e*, and *h*, respectively.

SUPPLEMENTAL APPENDIX

"Optical identification of hybrid magnetic and electric excitations in $Dy_3Fe_5O_{12}$ garnet" by

P. D. Rogers, Y. J. Choi, E. Standard, T. D. Kang, K. H. Ahn, A. Dubroka, P. Marsik, Ch. Wang, C. Bernhard, S. Park, S-W. Cheong, M. Kotelyanskii, and A. A. Sirenko

Expressions for the Adjusted Oscillator Strength (AOS) and the Adjusted Oscillator Strength Matching (AOSM) condition are developed for materials with $\mu \neq 1$. V. G. Veselago's results for semi-infinite magnetic materials [SA1, SA2] together with analytic expressions obtained by the authors are used in this treatment. It is assumed that magnetic and dielectric excitations can be modeled as Lorentzian oscillators. For a single hybrid mode, the dielectric and magnetic models are:

$$\varepsilon(\omega) = \varepsilon_\infty + \frac{S_e \omega_h^2}{(\omega_h^2 - \omega^2 - i\gamma\omega)},$$
$$\mu(\omega) = \mu_\infty + \frac{S_m \omega_h^2}{(\omega_h^2 - \omega^2 - i\gamma\omega)}, \tag{S1}$$

where $S_{e(m)}$ is the corresponding mode oscillator strength, $\gamma$ is the damping constant, and $\omega_h$ is the resonance frequency. Reflection from normally incident radiation is assumed throughout (AOI=0).

The semi-infinite case is examined first. Based on Veselago's work, it is assumed that the $s$ polarized reflection intensity $R_{ss}(\omega)$ is a function of $\sqrt{\varepsilon(\omega)/\mu(\omega)}$ [SA1,SA2]. Then, in the proximity of a single hybrid mode, for reflectivity:

$$R_{ss}(\omega) = f\left(\sqrt{\frac{\varepsilon(\omega)}{\mu(\omega)}}\right) = f\left(\sqrt{\frac{\varepsilon_\infty + \frac{S_e \omega_h^2}{(\omega_h^2 - \omega^2 - i\gamma\omega)}}{\mu_\infty + \frac{S_m \omega_h^2}{(\omega_h^2 - \omega^2 - i\gamma\omega)}}}\right) = f\left(\sqrt{\varepsilon_\infty \left(1 + \frac{\frac{S_e}{\varepsilon_\infty}\omega_h^2}{(\omega_h^2 - \omega^2 - i\gamma\omega)}\right) \bigg/ \mu_\infty\left(1 + \frac{\frac{S_m}{\mu_\infty}\omega_h^2}{(\omega_h^2 - \omega^2 - i\gamma\omega)}\right)}\right)$$

$$\approx f\left(\sqrt{\frac{\varepsilon_\infty}{\mu_\infty} + \frac{(\mu_\infty S_e - \varepsilon_\infty S_m)\omega_h^2}{\mu_\infty^2(\omega_h^2 - \omega^2 - i\gamma\omega)}}\right) = f\left(\sqrt{\frac{\varepsilon_\infty}{\mu_\infty} + \frac{S_R \omega_h^2}{(\omega_h^2 - \omega^2 - i\gamma\omega)}}\right),$$

(S2)

where $f(x) = (1-x)/(1+x)$. The AOSM condition, $S_m \varepsilon_\infty = S_e \mu_\infty$, is immediately apparent from Eq. (S2). Under this condition, the hybrid mode disappears from reflectivity and is a function of $\varepsilon_\infty$ and $\mu_\infty$ only: $R_{ss}(\omega) = f\left(\sqrt{\varepsilon_\infty / \mu_\infty}\right)|_{S_m \varepsilon_\infty = S_e \mu_\infty}$. In general, the hybrid resonance can be described with an AOS in reflection: $S_R = (\mu_\infty \cdot S_e - \varepsilon_\infty \cdot S_m)/\mu_\infty^2 \approx \mu_\infty \cdot S_e - \varepsilon_\infty \cdot S_m$. For a pure magnetic dipole at $\omega_h = \omega_m$, Eq. (S2) can be approximated for $S_e = 0$ and $\mu_\infty = 1$ as:

$$R_{ss}(\omega) = f\left(\sqrt{\varepsilon_\infty - \frac{\varepsilon_\infty S_m \omega_m^2}{(\omega_m^2(1+S_m) - \omega^2 - i\gamma\omega)}}\right) \approx f\left(\sqrt{\varepsilon_\infty - \frac{\varepsilon_\infty S_m \omega_m^2}{(\omega_m^2 - \omega^2 - i\gamma\omega)}}\right). \quad (S3)$$

The negative sign in Eq. (S3) corresponds to the inverted Lorentzian shape of a pure magnetic dipole with AOS: $S_R = S_m \cdot \varepsilon_\infty$. For hybrid modes, this inverted shape provides for the partial or complete cancellation of the electric and magnetic components at resonance. As is evident from Eq. (S3), a pole in the effective dielectric function measured, for example, in RAE experiments, is shifted from $\omega_m$, appearing at the longitudinal frequency $\omega_{LO} = \omega_m \cdot \sqrt{1+S_m}$. Note that this frequency shift is small due to $S_m \ll \mu_\infty$ for magnetic modes.

If light propagation in transmission is mainly driven by exponential decay and the extinction coefficient, according to Veselago, $T_{ss}(\omega)$ becomes a function of $\varepsilon(\omega) \cdot \mu(\omega)$:

$$T_{ss}(\omega) = F\left(\sqrt{\varepsilon(\omega) \cdot \mu(\omega)}\right) = F\left(\sqrt{\left(\varepsilon_\infty + \frac{S_e \omega_h^2}{(\omega_h^2 - \omega^2 - i\gamma\omega)}\right) \cdot \left(\mu_\infty + \frac{S_m \omega_h^2}{(\omega_h^2 - \omega^2 - i\gamma\omega)}\right)}\right)$$
$$= F\left(\sqrt{\varepsilon_\infty \cdot \mu_\infty + \frac{(S_e \cdot \mu_\infty + S_m \cdot \varepsilon_\infty) \cdot \omega_h^2}{(\omega_h^2 - \omega^2 - i\gamma\omega)} + \delta}\right) \approx F\left(\sqrt{\varepsilon_\infty + \frac{S_T \cdot \omega_h^2}{(\omega_h^2 - \omega^2 - i\gamma\omega)}}\right), \quad (S4)$$

where $F(y) = \exp(-j \cdot \omega \cdot d \cdot y)$. The AOS in transmission is: $S_T \approx S_e \cdot \mu_\infty + S_m \cdot \varepsilon_\infty$. Note that the two factors in $S_T$ are additive. The expressions for $S_R$ and $S_T$ allow for analysis of the interesting case of hybrid modes which cancel or disappear in reflectivity but remain strong in transmission.

A complete analysis of thin film reflectivity and transmission must involve the reflection from the backside of the sample, which depends on the thickness $d$. The opposing shapes of the Lorentzian profile of the magnetic and electric excitations motivate the calculation of $\frac{\partial R_{ss}(\omega_h)}{\partial \omega}$ and $\frac{dT_{ss}(\omega_h)}{d\omega}$. The two total derivatives require partial derivative expansion of the response functions as well as those of $r_{ss}$ and $t_{ss}$, the complex reflection and transmission coefficients. For a magnetic thin film whose principal axes are coincident with the laboratory system, $r_{ss}$ and $t_{ss}$ are given by [SA3]:

$$r_{ss} = \frac{q_{zs}\cos(q_{zs}d)(k_{z0}-k_{z2}) + i\left(\frac{q_{zs}^2}{\mu_{xx}} - k_{z0}k_{z2}\mu_{xx}\right)\sin(q_{zs}d)}{q_{zs}\cos(q_{zs}d)(k_{z0}+k_{z2}) - i\left(\frac{q_{zs}^2}{\mu_{xx}} + k_{z0}k_{z2}\mu_{xx}\right)\sin(q_{zs}d)},$$

$$t_{ss} = \frac{2k_{z0}q_{zs}}{q_{zs}\cos(q_{zs}d)(k_{z0}+k_{z2}) - i\left(\frac{q_{zs}^2}{\mu_{xx}} + k_{z0}k_{z2}\mu_{xx}\right)\sin(q_{zs}d)}$$

(S5)

where $k_{z0}$, $q_{zs}$ and $k_{z2}$ are the $z$ components of the wave vector in the incident, thin film and substrate media, respectively. At hybrid resonance, the following expressions for the two total derivatives are obtained:

$$\frac{dR_{ss}}{d\omega} \cong r_{ss}^* \cdot S_2 + r_{ss} \cdot S_2^* \quad \text{and} \quad \frac{dT_{ss}}{d\omega} \cong t_{ss}^* \cdot S_3 + t_{ss} \cdot S_3^* \tag{S6}$$

where $S_2$ and $S_3$ are given by:

$$S_2 = -\frac{2\omega_h}{\gamma_h^2} \frac{\alpha_e^{R_{TF}}(\omega_h)}{\sqrt{\mu(\omega_h)\varepsilon(\omega_h)}} \left(\mu(\omega_h)S_e + \varepsilon(\omega_h)S_m \frac{\alpha_m^{R_{TF}}(\omega_h)}{\alpha_e^{R_{TF}}(\omega_h)}\right)$$

$$S_3 = -\frac{2\omega_h}{\gamma_h^2} \frac{\alpha_e^{T}(\omega_h)}{\sqrt{\mu(\omega_h)\varepsilon(\omega_h)}} \left(\mu(\omega_h)S_e + \varepsilon(\omega_h)S_m \frac{\alpha_m^{T}(\omega_h)}{\alpha_e^{T}(\omega_h)}\right)$$

(S7)

The four $\alpha$ terms are components of the partial derivatives of the complex reflection and transmission coefficients taken with respect to the two response functions. Analytic solutions for these terms can be obtained starting from $r_{ss}$ and $t_{ss}$. For the $Dy_3Fe_5O_{12}$ material parameters with film thickness 0.55 mm, $\frac{\alpha_m^{R_{TF}}(\omega_h)}{\alpha_e^{R_{TF}}(\omega_h)}$ and $\frac{\alpha_m^{T}(\omega_h)}{\alpha_e^{T}(\omega_h)}$ are negative and positive, respectively, with absolute value equal to 1 (see FIG. S1). When these values are inserted into Eq. (S7), the upper and lower bracketed terms can be identified with the $S_R$ and $S_T$ terms discussed in the Veselago qualitative analysis above. These results are also consistent with the subtraction and addition of the AOS components in reflectivity and transmission, respectively.

The case where hybrid mode magnetic and electric dipole contributions completely cancel in reflection ($S_R = 0$) but add to $S_T$ in transmission requires the solution of the following simultaneous equation:

$$\mu(\omega_h) S_e + \varepsilon(\omega_h) S_m \frac{\alpha_m^{R_{TF}}(\omega_h)}{\alpha_e^{R_{TF}}(\omega_h)} = 0$$

$$\mu(\omega_h) S_e + \varepsilon(\omega_h) S_m \frac{\alpha_m^{T}(\omega_h)}{\alpha_e^{T}(\omega_h)} = S_T$$

(S8)

For the case of the fitted parameters for Dy-IG, $\frac{\alpha_m^{R_2}(\omega_h)}{\alpha_e^{R_2}(\omega_h)} \approx -1$, $\frac{\alpha_m^{T}(\omega_h)}{\alpha_e^{T}(\omega_h)} \approx 1$, $\mu(\omega_h) \approx 1$ and $\varepsilon(\omega_h) \approx \varepsilon_\infty$, Eq. (S8) has the approximate solution: $S_e \cong \frac{S_T}{2}$ and $S_m \cong \frac{S_T}{2\varepsilon_\infty}$.

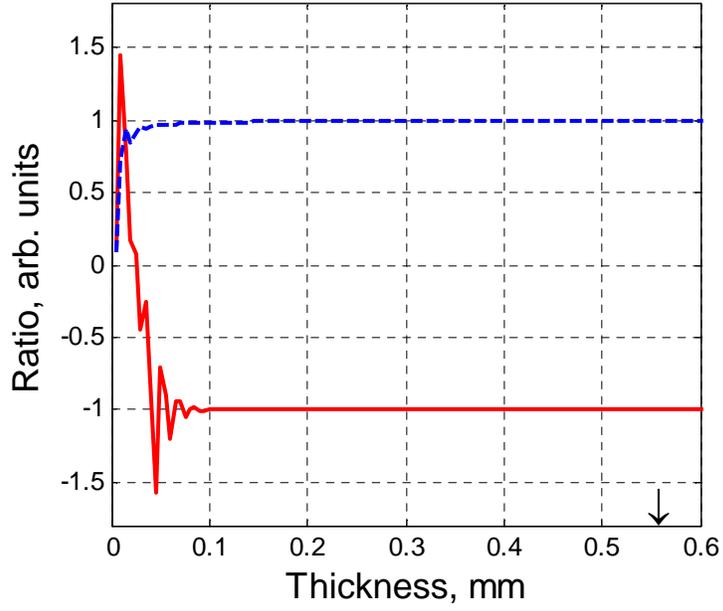

FIG. S1. Variability of the ratio of $\alpha$ terms with thin film thickness, $d$. $\varepsilon_\infty = 15.85$, $S_e = 0.100$, $S_m = 0.0063$ and $\omega_h = 78\,\text{cm}^{-1}$. $\dfrac{\alpha_m^{R_{TF}}(\omega_h)}{\alpha_e^{R_{TF}}(\omega_h)}$ is the bottom solid red line. $\dfrac{\alpha_m^{T}(\omega_h)}{\alpha_e^{T}(\omega_h)}$ is the top blue dashed line. For the $Dy_3Fe_5O_{12}$ sample with thickness $d = 0.55\,\text{mm}$, the opposite signs of these two ratios account for the subtraction of AOS contributions in reflectivity and the addition of the AOS contributions in transmission.